\documentclass[aps,pra,reprint,superscriptaddress,showpacs]{revtex4-1}
\usepackage{bm}
\usepackage{graphicx}
\begin{document}

\title{A method for simulating level anti-crossing spectra of diamond
crystals containing NV$^-$ color centers}

\author{S.V.~Anishchik}
\email[]{svan@kinetics.nsc.ru} \affiliation{Voevodsky Institute of
Chemical Kinetics and Combustion SB RAS, 630090, Novosibirsk,
Russia}

\author{K.L.~Ivanov}
\email[]{ivanov@tomo.nsc.ru} \affiliation{International Tomography
Center SB RAS, 630090, Novosibirsk, Russia}
\affiliation{Novosibirsk State University, 630090, Novosibirsk,
Russia}

\begin{abstract}
We propose an efficient method for calculating level anti-crossing
spectra (LAC spectra) of interacting paramagnetic defect centers
in crystals. By LAC spectra we mean the magnetic field dependence
of the photoluminescence intensity of paramagnetic color centers:
such field dependences often exhibit sharp features, such as peaks
or dips, originating from LACs in the spin system. Our approach
takes into account the electronic Zeeman interaction with the
external magnetic field, dipole-dipole interaction of paramagnetic
centers, hyperfine coupling of paramagnetic defects to magnetic
nuclei and zero-field splitting. By using this method, we can not
only obtain the positions of lines in LAC spectra, but also
reproduce their shapes as well as the relative amplitudes of
different lines. As a striking example, we present a calculation
of LAC spectra in diamond crystals containing negatively charged
NV centers.
\end{abstract}


\maketitle

\section{Introduction}

Owing to their unique properties \cite{Doherty2013} negatively
charged nitrogen-vacancy centers, NV$^-$ centers, in diamond
crystals are promising quantum objects for various exciting
applications, such as, e.g., nano-sensing
\cite{Balasubramanian2008,Rondin2014,Schirhagl2014,Kucsko2013,
Neumann2013}, quantum information processing
\cite{Jelezko2004,Maurer2012,vanderSar2012,Dolde2013,Dolde2014},
imaging of biological processes
\cite{Hall2012,Barry2016,Steinert2013,Lovchinsky2016}. Experiments
with NV$^-$ centers exploit optically detected magnetic resonance,
allowing one to probe the spin dynamics of single color centers
\cite{Gruber1997,Suter2017}, i.e., of single quantum objects. Such
extraordinary sensitivity, along with extremely long spin memory
times and decoherence times \cite{Balasubramanian2009} make
feasible various challenging experiments. Importantly, such
properties are preserved even at room temperature
\cite{GurudevDutt2007,Neumann2008,Hanson2006}, which dramatically
extends the range of existing and potential applications of NV$^-$
centers. One of the key issues for such applications is
interaction of NV$^-$ centers with other defect centers of the
diamond lattice. Since such defect centers are often ``dark''
centers, information about such interactions can be deduced
indirectly from Level Anti-Crossng (LAC) spectra. Experimentally
such LAC spectra of diamond single crystals containing NV$^-$
centers can be studied by monitoring the magnetic field dependence
of the photoluminescence intensity, which contains sharp features
associated with LACs \cite{Hanson2006,VanOort1989,
Epstein2005,Rogers2008,Rogers2009,Lai2009,Armstrong2010,
Anishchik2015,Broadway2016,Hall2016,Wickenbrock2016,Anisimov2016,
Zheng2017,Anishchik2017,Akhmedzhanov2017}. To probe such ``LAC
lines'' no resonant microwave or radiofrequency pumping is
required.

\begin{figure}\includegraphics[width=0.3\textwidth]{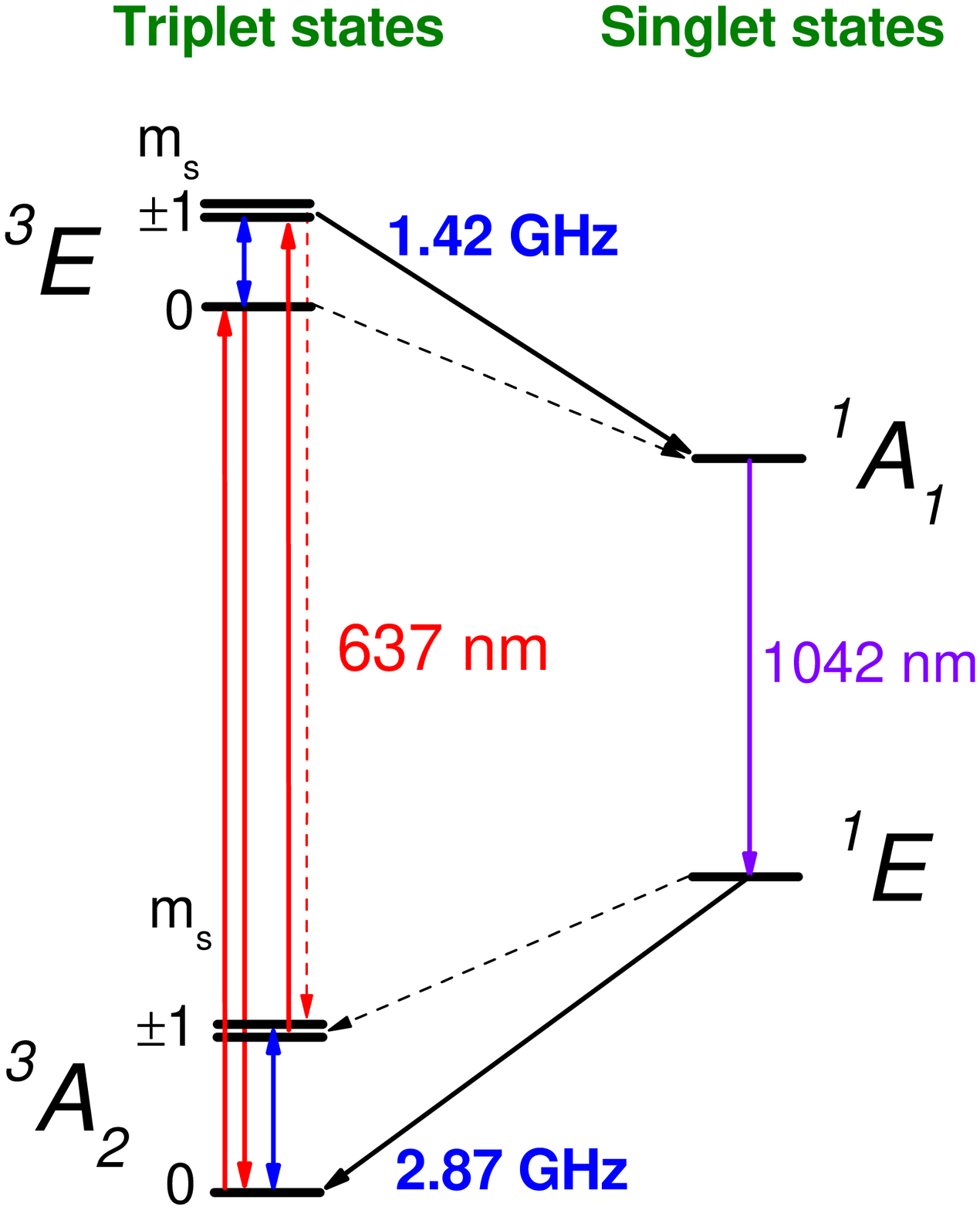}
\caption{Energy levels of an NV$^-$ center at zero magnetic field.
By arrows we show the zero-phonon line optical transitions,
inter-system crossing between the triplet and singlet states and
RF-induced transitions in the ground and excited states. Strong
transitions are shown by solid lines, weak transitions are shown
by dashed lines. States of the NV$^-$ center are classified
according to the spin multiplicity and irreducible
representations, $A$ and $E$, of the C$_{3v}$ point group. ZFS of
each triplet state is indicated, as well as the wavelengths of the
optical transitions. \label{shema}}\end{figure}

The scheme of the energy levels of an NV$^-$ center at zero
external magnetic field is shown  in Fig.~\ref{shema}, with
radiative and radiationless optical transitions and
radiofrequency-driven spin transitions indicated. The ground state
of the defect center is an electronic triplet state. Due to
zero-field splitting (ZFS) in the absence of a field the lowest
state in energy is the state with zero projection, denoted as
$M_s$, of the electron spin on the NV$^-$ center symmetry axis,
hereafter denoted as $\bm{r}_{NV}$. The excited state has a
similar structure of the spin energy levels, but with a smaller
ZFS value (by approximately a factor of two). For practical
applications of NV$^-$ centers, it is important that the ``optical
cycle'' gives rise to strong spin polarization
\cite{Delaney2010,Robledo2011,Goldman2015,Thiering2018}. The
mechanism of polarization formation is given by the dependence of
the inter-system crossing rates on the spin projection value
$M_s$.

Due to the peculiarities of the optical cycle, there is a strong
difference in the luminescence quantum yield upon excitation of an
NV$^-$ in the spin states with different $M_s$ values. When
excitation occurs from the $M_s=0$ state, the yield of subsequent
photoluminescence is close to unity. For this reason, the $M_s=0$
state is a ``bright'' state of the color center. By contrast,
after excitation from the $M_s=\pm1$ states the luminescence yield
is small due to the fast intersystem crossing process
$^3E\to{}^1A_1$; hence, these states are ``dark'' states. Owing to
the fact that the $^1E\to{}^3A_2$ process is also dependent on the
$M_s$ value (it is fast for the $M_s=0$ state); after light
irradiation the system goes to the $M_s=0$ state. This means that
the ground state of the NV$^-$ center acquires spin polarization
\cite{Loubser1977,Manson2006,Delaney2010}. Hence, NV$^-$ centers
can be spin-polarized by light excitation; such spin-polarized
states can also be studied by optically detected magnetic
resonance by using the remarkable property that NV$^-$ centers
have ``bright'' and ``dark'' spin states.

Spin dynamics of NV$^-$ centers can be thus probed via
photoluminescence, which also strongly depends on the magnetic
field. The reason is that at specific field strengths mixing of
the ``bright'' and ``dark'' states takes place, which gives rise
to a drop in the luminescence intensity. Specifically, such mixing
becomes efficient at LACs, which give rise to sharp features in
the field dependence of luminescence.

As usual, by a LAC we mean the following situation. Let us assume
that at a field strength $B_0=B_{LC}$ a level crossing occurs for
a pair of levels, $E_\mu$ and $E_\nu$, corresponding to the
eigen-states $|\mu\rangle$ and $|\nu\rangle$ of the main part of
the spin Hamiltonian $\hat{H}_0$. However, if there is also a
perturbation term to the full Hamiltonian, i.e.,
$\hat{H}=\hat{H}_0+\hat{V}$, such that
$V_{\mu\nu}=\langle\mu|\hat{V}|\nu\rangle\neq0$, the levels never
cross. In the mathematical sense this means that the $\mu$-th and
$\nu$-th eigen-states of $\hat{H}$ are different at any field; the
minimal splitting between them at $B_0=B_{LC}$ is equal to
$2|V_{\mu\nu}|\neq0$. In other words, the crossing is avoided: the
level crossing turns into a LAC. Importantly, at the LAC not only
the degeneracy of the energy levels is lifted, but also the
eigen-states change: the true eigen-states are no longer given by
$|\mu\rangle$ and $|\nu\rangle$ but by superposition states. As a
result, mixing of the $|\mu\rangle$ and $|\nu\rangle$ states takes
place. For this reason, LACs are of importance for analyzing the
magnetic field dependence of the photoluminescence intensity of
NV$^-$ centers: at a LAC involving a ``bright'' state and a
``dark'' state the intensity drops because initially
over-populated ``bright'' state becomes mixed with the ``dark''
state. Dips in the magnetic field dependence of the
photoluminescence intensity are thus associated with such LACs.
Therefore, we hereafter term this field dependence ``LAC
spectrum''.

Simulation of LAC spectra remains a challenging problem because of
the need to model the spin dynamics in a complex multi-level
system. As we show below, for simulating some LAC lines one should
treat a mutil-spin system, which dramatically increases the
dimensionality of the spin Hamiltonian matrices. In principle,
full quantum mechanical treatment of polarization transfer, which
would explicitly include the dynamics in both triplet states,
$^3A$ and $^3E$, is possible
\cite{Anishchik2015,Ivady2015,Anishchik2017,Sosnovsky2018};
however, it requires introducing the density matrix of a
multi-level system (including the energy levels of the excited
states) and working in the Liouville space rather than in the
Hilbert space. As a consequence, the dimensionality of the
Liouville super-matrices becomes too large and numerical solution
of the equation for the density matrix of the system becomes
extremely time consuming, if at all possible. In the present case,
this problem is further aggravated by the necessity to take into
account a second defect center, to which polarization is
transferred from the primarily polarized NV$^-$ center. For this
reason, we use simplifying assumptions, which allow us to treat
re-distribution of the light-induced spin polarization in a system
of two coupled defect centers having magnetic nuclei. The
simulation method proposed here is numerically efficient and
relatively easy to implement. As we show below, this approach
allows one to model LAC spectra, being able to reproduce not only the
positions of LAC lines (including weak satellites to main
lines) but also their relative amplitudes. Hence, we are able to
simulate LAC spectra, model LAC lines and assign them to
interaction of specific pairs of interacting defect centers. This
work is thus important for developing methods for detection of
``dark'' defect centers in diamond crystals.

\section{Experimental}

The experimental method is described in detail in a previous
publications \cite{Anishchik2015,Anishchik2017}. All experiments
were carried out using single crystals of a synthetic diamond. The
average concentration of NV$^-$ centers was $9.3\times10^{17}$
cm$^{-3}$. The samples were placed in a magnetic field, which is a
superposition of the constant field, $\bm{B}_0$, and a weak field
modulated with the amplitude $B_m$ at the frequency $f_m$, and
irradiated by the laser light at a wavelength of 532~nm (the
irradiation power was 400 mW). The beam direction was
perpendicular to the magnetic field vector $\bm{B}_0$. The laser
light was linearly polarized and the electric field vector
$\bm{E}$ was perpendicular to $\bm{B}_0$. In experiments, we
precisely oriented the sample such that the magnetic field was
parallel to the $[111]$ crystal axis.

The luminescence intensity was measured by a photo-multiplier. The
resulting signal was sent to the input of a lock-in detector to
increase the detection sensitivity. Such a method indeed allows
one to enhance the sensitivity and to resolve multiple LAC lines.
Due to lock-in detection, lines in experimental LAC spectra have
dispersive shapes. For this reason, we also integrate the obtained
LAC spectra and compare them to calculated spectra. The modulation
frequency $f_m$ was 17~Hz and modulation amplitude was
${B}_m=0.5$~G. All experiments were carried out at room
temperature.

\begin{figure}
   \includegraphics[width=0.4\textwidth]{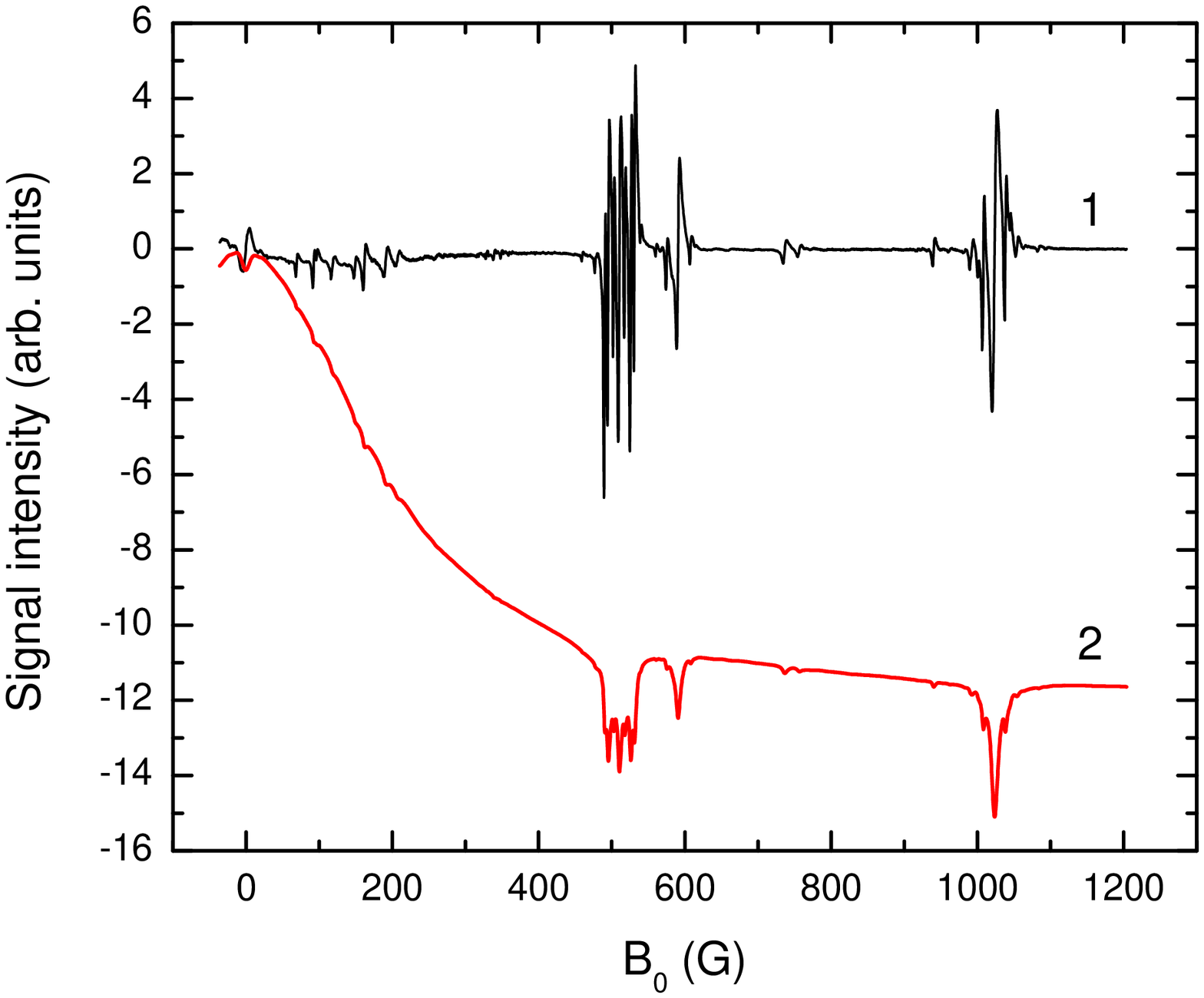}
   \caption{Experimental LAC spectrum of a diamond
   single crystal containing NV$^-$ centers. Here we show the signal
   obtained by using
   lock-in detection (curve 1) with the polarization of light $\bm{E}
   \bot \bm{B}_0$ and the integrated spectrum (curve 2). The modulation
   frequency is 17~Hz, the modulation amplitude is 0.5~G.
   \label{exper}}
\end{figure}

In this work, we present the experimental results, namely, LAC
spectra, obtained earlier \cite{Anishchik2017}. In
Fig.~\ref{exper} we present the LAC spectrum of the sample
obtained by using lock-in detection of photoluminescence and also
the integrated spectrum. Both spectra contain multiple sharp LAC
lines. Below, we compare parts of the integrated spectrum with
calculation results. We discuss only some of the LAC lines and do
not model the lines, which are not yet assigned (these are the
lines found in the range 50-250~G and lines at about 750~G and
950~G).

\section{Theory}

\subsection{Spin Hamiltonian}

Generally, the spin Hamiltonian  $\hat{H}$ of two
interacting paramagnetic centers in the external $\bm{B}_0$ field
can be written as a sum of the main Hamiltonian and a perturbation
term:
\begin{equation}
\hat{H} =\hat{H}_0+\hat{V}\label{h0v} \label{eq1}
\end{equation}
where
\begin{equation}
\hat{H}_0 =\beta\bm{B}_0 {\cal
G}_1\bm{\hat{S}}_1+\bm{\hat{S}}_1{{\cal D}}_1\bm{\hat{S}}_1+
\beta\bm{B}_0{\cal G}_2\bm{\hat{S}}_2+\bm{\hat{S}}_2{{\cal
D}}_2\bm{\hat{S}}_2\label{eq2}
\end{equation}
 and
\begin{eqnarray}
\nonumber \hat{V} =&\sum\limits_i \bm{\hat{S}}_1{\cal
A}_i\bm{\hat{I}}_i+
 \sum\limits_i \bm{\hat{I}}_i{\cal Q}_i\bm{\hat{I}}_i
+\sum\limits_j \bm{\hat{S}}_2{\cal A}_j\bm{\hat{I}}_j+
\sum\limits_j \bm{\hat{I}}_i{\cal Q}_j\bm{\hat{I}}_j\\
~&~~+{D}_{dd}\left[3{(\bm{\hat{S}}_1\cdot\bm{n}_{12})
(\bm{\hat{S}}_2\cdot\bm{n}_{12})}-({\bm{\hat{S}}_1\cdot\bm{\hat{S}}_2})\right],
\label{eq3}
 \end{eqnarray}
Here $\hat{\bm{S}}_1$ and $\hat{\bm{I}}_i$ are the electron and
nuclear spin operators of the NV$^-$ center, $\hat{\bm{S}}_2$ and
$\hat{\bm{I}}_j$ correspond to the other defect center. The main
part of the Hamiltonian takes into account the electronic Zeeman
interaction (described by the $g$-tensors ${\cal G}_1$ and ${\cal
G}_2$ with $\beta$ being the Bohr magneton) and ZFS (given by the
ZFS  tensors ${\cal D}_1$ and ${\cal D}_2$). The perturbation term
includes the hyperfine couplings (HFCs) to magnetic nuclei (given
by the HFC tensors ${\cal A}_i$ and ${\cal A}_j$), nuclear
quadrupolar couplings (given by the tensors ${\cal Q}_i$ and
${\cal Q}_j$) and the electronic dipole-dipole interaction between
the two centers, which depends on the distance between them:
${D}_{dd}\propto r^{-3}_{12}$, where $\bm{r}_{12}$ is the vector
connecting the two defect centers,
$\bm{n}_{12}=\bm{r}_{12}/r_{12}$ is unity length vector parallel
to $\bm{r}_{12}$. Here, for simplicity, nuclear Zeeman interaction
is neglected as we are working at relatively low magnetic fields.
In this work, we will specify the Hamiltonian and all relavant
parameters (HFC and quadrupolar couplings, ZFS values, ${D}_{dd}$)
in the units of Hz. In this work, all tensors are denoted by
calligraphic capital letters.

The interaction  terms in eqs. (\ref{eq2}) and (\ref{eq3}) come
from anisotropic interactions and therefore depend on the
molecular orientation and on the choice of the coordinate frame.
The individual terms of the Hamiltonian become simple, when when
the reference frame coincides with the Principal Axes System (PAS)
of the corresponding interaction tensor: in such a frame the
tensor simply becomes diagonal and its non-zero elements are the
principle values of the tensor. As a consequence, the relevant
terms in their PASs are written as follows (here $k=1,2$):
\begin{eqnarray}
\nonumber
  \beta\bm{B}_0 {\cal G}_k\bm{\hat{S}}_k&=&
  \beta\left[g_{k,||}B_{0,z}\hat{S}_{kz}+g_{k,\perp}B_{0,x}\hat{S}_{kx} \right. \\
  ~~ &~& \left. +g_{k,\perp}B_{0,y}\hat{S}_{ky}\right]\label{pas1}
\end{eqnarray}
\begin{equation}
\bm{\hat{S}}_k{{\cal
D}}_k\bm{\hat{S}}_k=D_k\left[\hat{S}_{kz}^2-\frac{2}{3}\right]\label{pas2}
\end{equation}
\begin{equation}
\bm{\hat{S}}_k{\cal
A}_i\bm{\hat{I}}_i=A_{i,||}\hat{S}_{kz}\hat{I}_{iz}+A_{i,\perp}\hat{S}_{kx}\hat{I}_{ix}
+A_{i,\perp}\hat{S}_{ky}\hat{I}_{ky}\label{pas3}
\end{equation}
\begin{equation}
\bm{\hat{I}}_i{\cal
Q}_i\bm{\hat{I}}_i=Q_i\left[\hat{I}_{iz}^2-\frac{1}{3}I_i(I_i+1)\right].\label{pas4}
\end{equation}
Here we take into account that all defect centers considered here
have axial symmetry; consequcntly, the $x,y$-components of all
tensors are identical (denoted by $\perp$ instead of $x,y$),
generally, being different from the $z$-component (denoted by
$||$). Hence, $g_{i,||},g_{i,\perp}$ are the relevant components
of the ${\cal G}_k$ tensor; $D_k$ is the relevant component of the
ZFS tensor (hereafter we always assume that $E_k=0$, i.e., due to
axial symmetry the $x$ and $y$ directions are equivalent);
$A_{i,||},A_{i,\perp}$ are the components of the ${\cal A}_i$
tensor; $Q_i$ is the quadrupolar coupling of the corresponding
nucleus. In any frame different from PAS additional terms appear
in the interaction tensor and, consequently, in the Hamiltonian.
To define frame rotations one can use different methods, e.g., one
can specify the three  angles for consecutive Euler rotations.
Here we do not dwell into details of frame rotations, which are
done in a standard way \cite{Mehring1976}.

In this work, we always choose the reference frame such that the
$z$-axis is parallel to the $\bm{r}_{NV}$ vector of the first
defect center (which is always an NV$^-$ center). This means that
the tensors ${\cal G}_1$, ${\cal D}_1$, ${\cal A}_i$, ${\cal Q}_i$
have the simple form given by eqs. (\ref{pas1})-(\ref{pas4}). In
this situation, if the $\bm{B}_0$ vector is parallel to the
$[111]$ crystal axis, the magnetic field can be either parallel to
$\bm{r}_{NV}$, so that only the $B_{0,z}$ component is non-zero,
or tilted by $\theta_t=109.47^\circ$ (tetrahedral angle), so that
the transverse field components are present as well. For the
second center we assume that it (i) has the same orientation
meaning that the relevant interaction tensors also take the simple
form, expected for the PAS or (ii) its orientation is tilted by
$\theta_t$ with respect to  $\bm{r}_{NV}$.

\begin{figure}
   \includegraphics[width=0.4\textwidth]{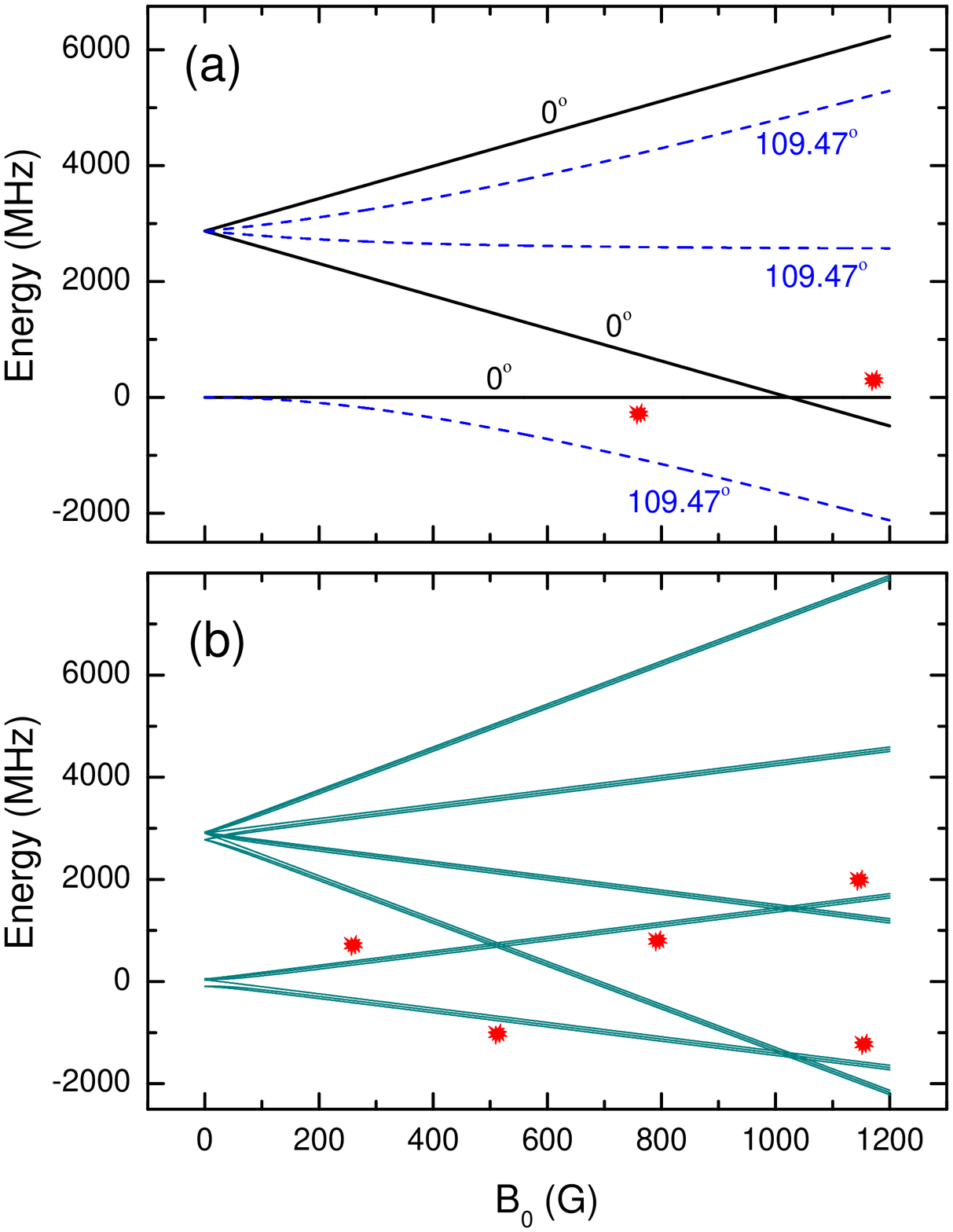}
   \caption{Calculated energy levels of the ground state of an isolated
   NV$^-$ center (a) and the system of NV$^-$ and
   P1 centers (b) in an external magnetic field $B_0$. Stars highlight
   the ``bright'' energy levels of the NV$^-$ center.
   \label{enlevels}}
\end{figure}

In Fig.~\ref{enlevels}(a) we present the energy levels of a single
isolated NV$^-$ center for the two possible orientations with
respect to the external magnetic field, which are existing when
the [111] crystal axis is parallel to the $\bm{B}_0$ field.
Specifically, there is an orientation where $\bm{r}_{NV}$ is
parallel to $\bm{B}_0$ (1/4 of all centers have this orientation)
and an orientation where the angle between $\bm{r}_{NV}$ and
$\bm{B}_0$ is equal to $\theta_t$ (3/4 of all centers have this
orientation). In Fig.~\ref{enlevels} we show the energy levels for
both possible orientations; in the calculation we take into
account the Zeeman interaction, ZFS, HFC of the electron spin of
the NV$^-$ center with the spin of the nitrogen nucleus and
nuclear quadrupolar coupling. When the angle between $\bm{r}_{NV}$
and $\bm{B}_0$ is equal to $\theta_t$ there are no level crossings
in the ground state. However, once $\bm{r}_{NV}$ is parallel to
$\bm{B}_0$, at the field of $B_0=D=1024$~G there is a crossing
between the ``bright'' $M_s=0$ level and the ``dark'' $M_s=-1$
level. HFC turns this crossing into a LAC; spin mixing at this LAC
brings the system from the $M_s=0$ state to the $M_s=-1$ state.
Consequently, the population of the ``bright'' state decreases,
giving rise to a reduction of the photoluminescence intensity at
this field. As a result, a sharp LAC line appears in the field
dependence of the luminescence intensity, see Fig.~\ref{exper}.

In Fig.~\ref{enlevels}(b) as a representative example we present
the calculated energy levels of a coupled system comprising an
NV$^-$ center and a P1 center. P1 center is a lattice defect in
diamond, in which the carbon atom is replaced by a neutral
nitrogen atom. P1 center is paramagnetic having the electron spin
of 1/2; symmetry properties of this defect center are described by
the C$_{3v}$ group, same as for NV$^-$ center. The total number of
states in the considered system of two paramagnetic centers (each
having a $^{14}$N spin-1 nucleus) is equal to
$54=3\times3\times2\times3$. The interaction between the defect
centers is the magnetic dipole-dipole interaction. In the
calculation we assume that the symmetry axes of both centers are
parallel to the $\bm{B}_0$ vector. Since the energy levels are
plotted for the total energy of the system of two defect centers,
the energy of the $M_s=0$ level of the NV$^-$ center is no longer
constant, as in Fig.~\ref{enlevels}(a), but depends on the ${B}_0$
field due to the Zeeman interaction of the P1 center with the
external field. One can see that in the case under consideration
there are multiple level crossings. HFCs and electronic
dipole-dipole interaction turn these crossings into LACs. As
previously, LACs of the $M_s=0$ levels (highlighted with asterisks
in the Figure) with the $M_s=-1$ levels give rise to a decrease of
the population of the ``bright'' state; consequently, sharp LAC
lines appear in the spectrum. In the present case, in addition to
the lines at around 1024~G more lines appear at around $B_0\approx
D/2\approx500$~G, which are traditionally called
``cross-relaxation lines''. Below we demonstrate clearly that
these lines are due to the coherent polarization transfer between
paramagnetic defect centers.

The analysis of the energy level diagram, like the one shown in
Fig.~\ref{enlevels}, allows one to determine the level crossing
points. This is, however, not sufficient for analyzing LAC spectra
because one also needs to know the efficiency of polarization
transfer between the ``bright'' and ``dark'' states at each
crossing. Hence, for quantitative analysis of LAC spectra the
theoretical treatment needs to be extended.

\subsection{Simulation method}

To calculate LAC spectra, we adapted an approach developed
previously to describe coherent polarization transfer in coupled multi-spin
systems \cite{Ivanov2008}. In order to use this approach, we also
introduce the following simplifications:

\begin{enumerate}
    \item We assume that photo-excitation of the NV$^-$ center is a
    very fast process as compared to the spin dynamics of hyperpolarized
    NV$^-$ centers in the ground state. The same holds for
    the inter-system crossing processes, ${^3E} \to {^1A_1}$ and
    ${^1E} \to {^3A_2} $, and fluorescence process,
    ${^3E} \to {^3A_2}$. Hence, we assume that all photo-induced
    processes     do not affect polarization transfer and only provide the initial
    spin polarization of the NV$^-$ center.

    \item We consider only the spin evolution of the ground state
    of the system: the reason is that the system spends only a small
    fraction of time in the excited state. We also do not see any traces of
    excited-state LACs in the measured spectrum shown in Fig.~\ref{exper}.
    We also assume that     the spin Hamiltonian of the NV$^-$ center interacting
    with another defect center does not depend on time.

    \item We completely neglect spin relaxation. Re-distribution of
    the polarization in the NV$^-$ center, as well as between the
    NV$^-$ center and another defect center in the crystal lattice
    is thus treated as a coherent process in the same manner as before
    \cite{Ivanov2008,Ivanov2006}.
\end{enumerate}

We introduce the initial spin density matrix of the system in the
form of the direct product (Kronecker product) of the individual
electronic and nuclear spin density matrices:
 \begin{equation}
\rho_0=\rho_{S1} \otimes \rho_{S2}\otimes \rho_{I1}\otimes ...
\otimes \rho_{In}.\label{instate}
 \end{equation}
Here $\rho_{S1}$ is the electron spin density matrix of the NV$^-$
center is a completely or partially  polarized  state
  \begin{equation}
\rho_{S1}=\alpha\left(%
\begin{array}{ccc}
  0 & 0 & 0 \\
  0 & 1 & 0 \\
  0 & 0 & 0 \\
\end{array}%
 \right)
 +\frac{1-\alpha}{3}
 \left(%
\begin{array}{ccc}
  1 & 0 & 0 \\
  0 & 1 & 0 \\
  0 & 0 & 1 \\
\end{array}%
 \right),\label{alpolar}
  \end{equation}
where $\alpha$ is the degree of the light-induced polarization of
the $^3A$ triplet state of the NV$^-$ center. The other matrices,
$\rho_{S2}$ (the electron spin density matrix of the second defect
center) and $\rho_{Ii}$ (the nuclear spin density matrices of the
NV$^-$ center) and $\rho_{Ij}$ (the nuclear spin density matrices
of the second defect center) are the equilibrium density matrices
of the corresponding spin systems. Hence, if we neglect the tiny
thermal spin polarization, they are equal to $N^{-1}\hat{E}$ where
$\hat{E}$ is the unity matrix of the corresponding dimensionality
$N$ (here the coefficient $N^{-1}$ is introduced to provide the
normalization condition ${\rm Tr}\{\rho_0\}=1$).

The state described by eq. (\ref{instate}) is a non-stationary
state (i.e., it is not an eigen-state) of the Hamiltonian
(\ref{eq1}); therefore, the density matrix evolves with time. The
spin evolution can be described quantitatively in the simplest way
in the eigen-basis of the $\hat{H}$ operator. In this basis, the
initial state is as follows:
 \begin{equation}
 \rho^{eb}_0=\hat{V}^{-1}\rho_0\hat{V},
 \end{equation}
where $\hat{V}$ is the matrix composed of the eigen-vectors of
$\hat{H}$. Since the Hamiltonian $\hat{H}$ is time-independent,
the matrices $\hat{V}$ and $\hat{V}^{-1}$ are the same at any
instant of time. In the new basis the Hamiltonian is diagonal and
has the following matrix elements:
$\hat{H}^{eb}_{ij}=E_i\delta_{ij}$, where $E_i$ is the $i$-the
eigen-value of the Hamiltonian (i.e., the energy of the $i$-th
state), $\delta_{ij}$ is the Kronecker delta. As we express the
energy in the frequency units, the density matrix obeys the
following equation:
 \begin{eqnarray}
 \nonumber i\frac{d\rho^{eb}_{ij}}{dt}&=\left[\hat{H}^{eb},\rho^{eb}\right]=
2\pi\sum_k \left(E_i \delta_{ik}\rho_{kj}^{eb}-\rho_{ik}^{eb}E_k
\delta_{kj}
\right)\\
~~~&=2\pi\left(E_i-E_j\right)\rho_{ij}^{eb}.
 \end{eqnarray}
 Consequently, we can obtain:
  \begin{equation}
 \rho^{eb}_{ij}(t)=\left\{\rho^{eb}_{0}\right\}_{ij}\exp(-2 \pi i
 \Delta_{ij} t),
 \end{equation}
 where $\Delta_{ij}=E_i-E_j$.

By going back to the original basis of spin states, we obtain the
following expression for the time-dependent density matrix:
 \begin{equation}
 \rho(t)=\hat{V}\rho^{eb}(t)\hat{V}^{-1},
 \end{equation}
which allows us to calculate the population of the ``bright''
state (the $M_s=0$ spin state of the $^3A$ state of the NV$^-$
center) at any instant of time:
\begin{equation}
\rho_{00}(t) ={\rm Tr}\{\hat{P}_0\rho(t)\},\label{r00}
 \end{equation}
where $\hat{P}_0=|0\rangle\langle0|$ is the projector on the
$M_s=0$ state, $|0\rangle$.

The coherent spin evolution described by eq. (\ref{r00}) is
interrupted at the instant of time $t$ after light excitation,
i.e., after the NV$^-$ center in the ground state absorbs a
photon. After that, the system goes back into the initial state,
see eq. (\ref{instate}), emitting a photon. If the light intensity
is constant, the distribution of the coherent spin evolution times
is exponential $\exp (-t/\tau)/\tau$, with the mean evolution
time, $\tau$, depending on the light intensity and the probability
of light absorption by the NV$^-$ center.

To evaluate the photoluminescence intensty we need to average the
$M_s=0$ population, $\rho_{00}(t)$, over instants of time when the
photon is absorbed. This averaged population is as follows:
\begin{eqnarray}
 \nonumber\langle\rho_{00}\rangle&=\langle\rho_{00}(t)\rangle=
 \displaystyle\frac{1}{\tau}\displaystyle\int_0^\infty
\rho_{00}(t)\exp(-t/\tau)dt\\
~~~~&={\rm Tr}\{\hat{P}_0\hat{V}\rho^{st}\hat{V}^{-1}\},
 \end{eqnarray}
where
\begin{eqnarray}
 \nonumber \rho^{st}_{ij}=& \langle \rho^{eb}(t) \rangle_{ij}=
 \rho_{0,ij}^{eb}\displaystyle\frac{1}{\tau}\displaystyle\int_0^\infty \exp(-t/\tau -
 2 \pi i \Delta_{ij} t)dt \\
 ~~~~&=\rho_{0,ij}^{eb}/(1+2 \pi i \Delta_{ij} \tau).
 \end{eqnarray}
We would like to note that here averaging is performed in the
eigen-basis, which makes the calculation much simpler. This is
possible because all transformations used here are linear
transformations and also due to the fact that the matrices
$\hat{V}$ and $\hat{V}^{-1}$ are time-independent.

\section{Results}

\subsection{LAC spectra of different systems}

In this subsection, we present the calculated LAC spectra for isolated
NV$^-$ centers as well as for pairs of interacting defect centers.
Such calculations allow us to describe different groups of lines
found in the experimental LAC spectrum. We would like to emphasize
that each calculation can reproduce only some of the observed LAC
lines. The reason is that in a real experiment, different NV$^-$
centers have different environment. For this reason, the
experimental LAC spectrum is a superposition (with appropriate
weights) of the spectra for isolated NV$^-$ centers as well as
for systems comprising an NV$^-$ center and another paramagnetic
center. In addition, in the proximity of the paramagnetic centers
there can also be $^{13}$C nuclei, which affect LACs and become
manifest in the observed LAS spectra. Here we present calculations
for isolated NV$^-$ centers and for NV$^-$ centers interacting
with other defect centers; such calculations will be done
neglecting HFCs to $^{13}$C nuclei and also taking such HFCs into
account in order to describe weak satellites to main LAC lines.
For the sake of simplicity, in this subsection we always assume a
significantly long $\tau$; the effect of the evolution time on the
LAC spectra is considered in a separate subsection.

In this work, we always evaluate the photoluminescence signal as
the averaged population, $\langle\rho_{00}\rangle$, of the $M_s=0$
state of the NV$^-$. Hence, we term the $\langle\rho_{00}\rangle$
dependence on the $B_0$ magnetic field as ``LAC spectrum'',
assuming that the photoluminescence intensity is directly
proportional to $\rho_{00}$.

\begin{figure}
   \includegraphics[width=0.4\textwidth]{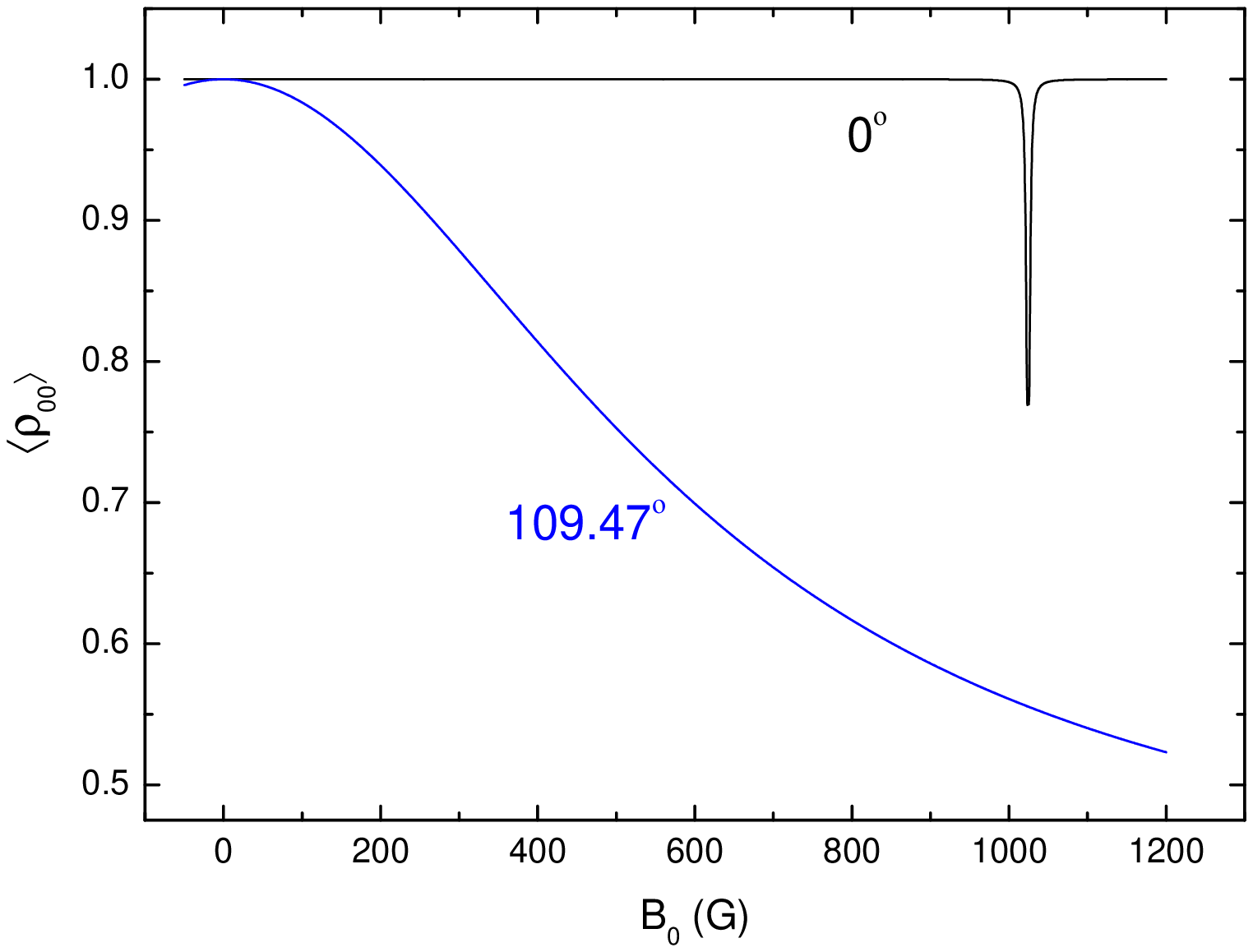}
   \caption{LAC spectrum of an isolated NV$^-$ center for the case
   $\bm{r}_{NV}||\bm{B}_0$ and for the case where the angle between
   $\bm{r}_{NV}$ and $\bm{B}_0$ is equal $\theta_t$.
    \label{NVlong}}
\end{figure}

In Fig.~\ref{NVlong} we show the calculated LAC
spectrum for a single NV$^-$ center. In the calculation we used
the Hamiltonian $\hat{H}$ of the form:
\begin{eqnarray}
\nonumber \hat{H}&=\beta\left[g_{||}B_{0,z}\hat{S}_z+g_\perp
(B_{0,x}\hat{S}_x+B_{0,y}\hat{S}_y)\right]+\\
\nonumber~&+D\left[\hat{S}^{2}_{z}-\frac{2}{3}\right]
+A_{||}\hat{S}_z\hat{I}_z+A_\perp\left[\hat{S}_x\hat{I}_x+\hat{S}_y\hat{I}_y\right]\\
&+P\left[\hat{I}_{z}^2-\frac{1}{3}I(I+1)\right].
 \label{hsingle}
\end{eqnarray}
Hereafter, in calculations we use the following magnetic
parameters for the NV$^-$ center \cite{Yavkin2016}: $g_{||}=2.0029$,
$g_{\perp}=2.0031$, $D_1=2872$~MHz, $E=0$, $A_{||}=-2.2$~MHz,
$A_{\perp}=-2.7$~MHz, $Q=-4.8$~MHz.

When the symmetry axis of the NV$^-$ center is parallel to
external field vector $\bm{B}_0$, there is a single sharp LAC line
at the field of 1024~G, where the $M_s=0$ and $M_s=-1$ levels
cross, as shown in Fig.\ref{enlevels}. At this LAC, the HFC to the
$^{14}$N nucleus of the defect center becomes active, turning the
level crossing into a LAC. Due to spin mixing at this LAC, the
population of the ``bright'' state is reduced and a dip in the
luminescence intensity is found. When the angle between the
$\bm{r}_{NV}$ and $\bm{B}_0$ vectors is equal to $\theta_t$ there
are no level crossings (see Fig.\ref{enlevels}); consequently,
there are no sharp lines in the spectrum. Instead, there is a
smooth decay of the population of the $M_s=0$ state upon increase
of the field strength. A similar decrease of the photoluminescence
can be observed experimentally, as can be seen in Fig.\ref{exper}.
The reason for this effect is that the field component, which is
perpendicular to the NV$^-$ symmetry axis, gives rise to
transitions between the triplet sublevels of the ground state. As
a result, the ``bright'' $M_s=0$ state is mixed with the ``dark''
$M_s=\pm1$ states and its population decreases giving rise to the
decrease of the photoluminescence intensity. Hence, the
calculation for an isolated NV$^-$ center can describe the smooth
decay of the luminescence intensity with increasing magnetic field
strength and also the LAC line at 1024~G but not other LAC lines
found in experiments. To describe these lines, we need to perform
calculations for two interacting defect centers with one of them
being the NV$^-$ center.

\begin{figure}
   \includegraphics[width=0.4\textwidth]{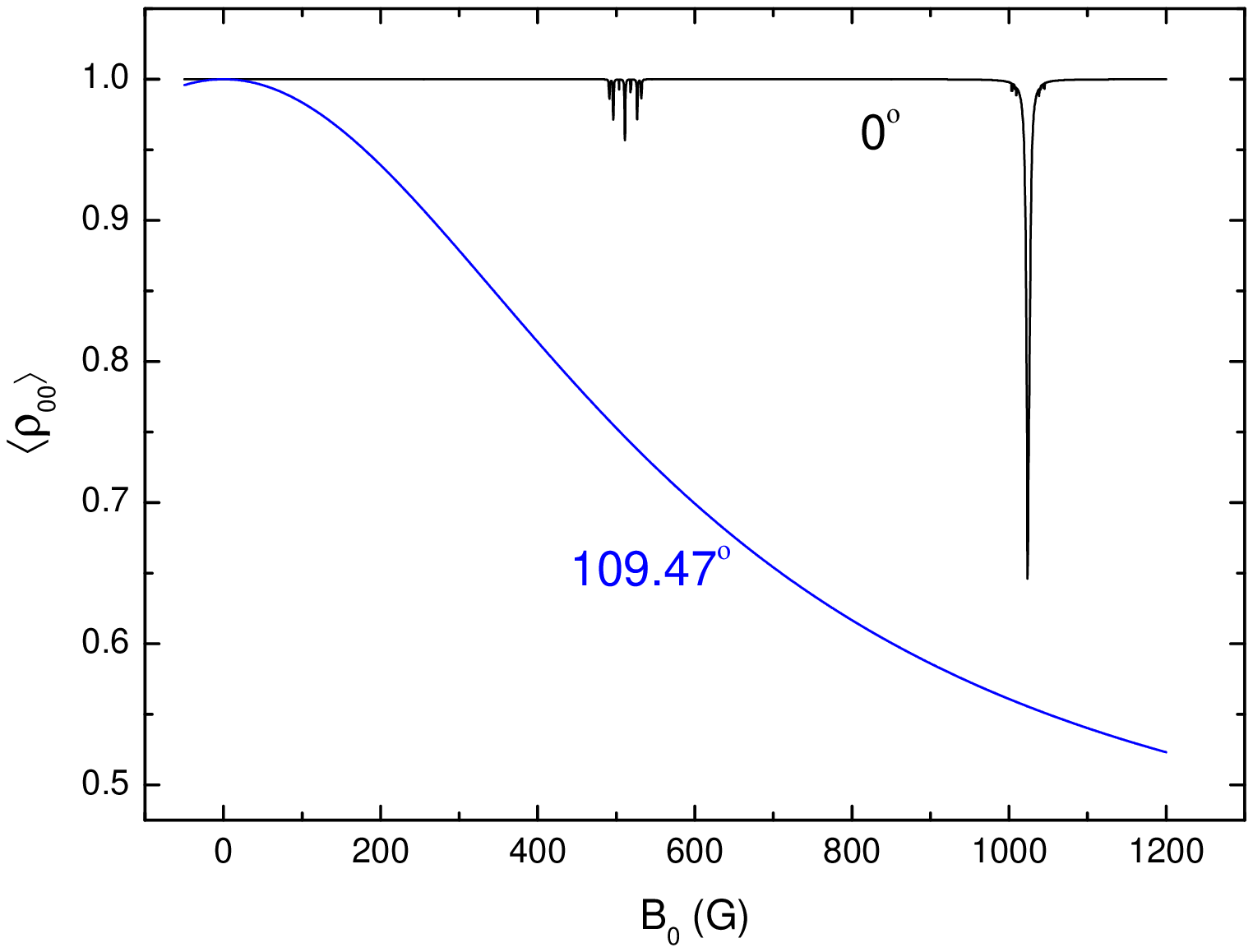}
   \caption{LAC spectrum of interacting NV$^-$ and P1 centers for
   the orientation where the NV$^-$ axis is parallel to the field
   or tilted by $\theta_t$. In the calculation, we have performed
   averaging over all four possible orientations of the P1 center.
   \label{P1long}}
\end{figure}

In Fig.~\ref{P1long} we show the calculated LAC spectrum for the
system of two interacting defect centers, NV$^-$ center and P1
center. In the calculation, we make use of the Hamiltonian given
by eqs. (\ref{eq1}-\ref{eq3}), where $S_1$, $I_1$, ${\cal
G}_1$, ${\cal A}_1$, ${\cal Q}_1$ being the parameters of the
NV$^-$ center, whereas $S_2$, $I_2$, ${\cal G}_2$, ${\cal A}_2$,
${\cal Q}_2$ are the parameters of the P1 center. In the
calculation, we use the following parameters:
$g^2_{||}=g^2_{\perp}=2.0023$, $A^2_{||}=114$~MHz,
$A^2_{\perp}=81$~MHz, $Q_2=Q_1=-4.8$~MHz, ${D}_{dd} =
1$~MHz. Hence, the parameters for the NV$^-$ center are the same
as those in the calculation shown in Fig.~\ref{NVlong}. All
calculations are performed using averaging over the four possible
orientations of the P1 center.

When the angle between the $\bm{B}_0$ and $\bm{r}_{NV}$ vectors is
equal to $\theta_t$, like in the previous case, there is a smooth
decrease of $\langle\rho_{00}\rangle$ upon increasing the magnetic
field strength and no sharp lines are seen. The reason for this is
exactly the same as in the previous case of an isolated NV$^-$
center: there is mixing of the triplet sublevels by the
perpendicular field component and there are no level crossings.

In the case $\bm{r}_{NV}||\bm{B}_0$, in the LAC spectrum there are
two groups of sharp lines seen at around 1024~G and at around
500~G. These lines are also discussed below in further detail.
They are caused by the level crossings shown in in
Fig.\ref{enlevels}; however, we would like to stress that in the
calculation presented in Fig.~\ref{P1long} all four possible
orientations of the P1 center are taken into account. The fact
that the two groups of lines are found at the magnetic fields of
about $D_1$ and $D_1/2$ is due to the very small difference in the
$g$-factors of the NV$^-$ center and P1 center. Hence, the present
calculation can account for the LAC lines at around 500~G.

\begin{figure}
   \includegraphics[width=0.4\textwidth]{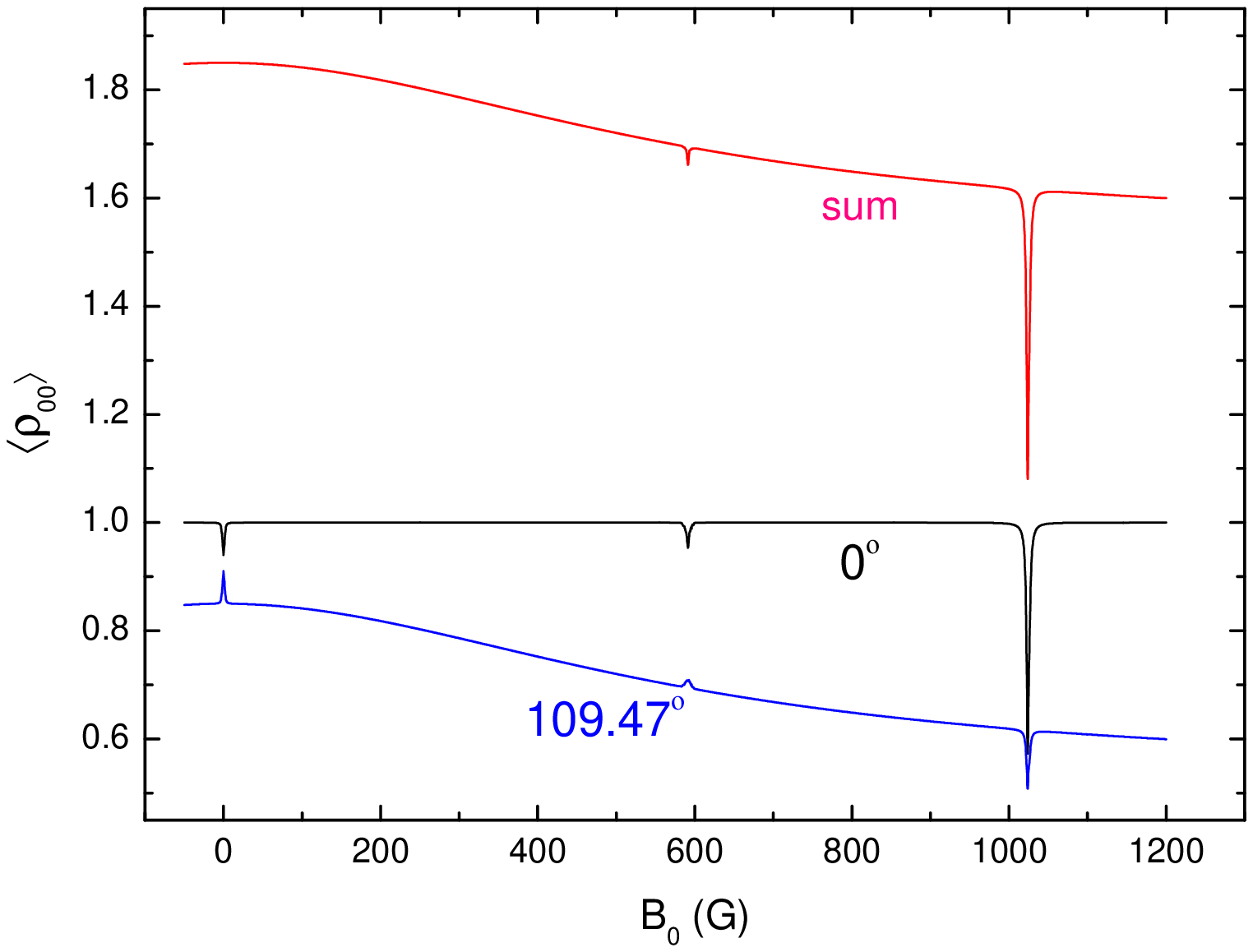}
   \caption{LAC spectrum of a system of two interacting NV$^-$ centers;
   one of them is oriented parallel to the magnetic field and the other one
   is oriented at the $\theta_t$ angle to the field (averaging over the
   three possible orientations of the second center is performed). The
   degree of polarization $\alpha$ (see eq. (\ref{alpolar})) for the first
   NV$^-$ center is equal to 1; for the second center it is 0.7. In the
   plot we show the average population $\langle\rho_{00}\rangle$ of the M$_s=0$
   state for each NV$^-$ center as well as the sum of these populations.
   \label{NVNVlong}}
\end{figure}

In Fig.~\ref{NVNVlong} we show the calculation result for a pair
of NV$^-$ centers, where one of them is oriented parallel to
$\bm{B}_0$, while the other one is oriented at the $\theta_t$
angle to the magnetic field. In the calculation, averaging over
three possible orientation of the second NV$^-$ center is
performed. Parameters of the calculation are the same as those in
previous calculations. For calculating the LAC spectra, we use
different degree of spin polarization $\alpha$ of the NV$^-$
centers having different orientations with respect to the magnetic
field. The reason is that the probability of absorbing a photon by
an NV$^-$ center depends on the polarization of the incident
light. Specifically, the absorption probability is maximal when
the polarization vector is perpendicular to $\bm{r}_{NV}$ being
close to zero when the two vectors are parallel. In our
experiments, the polarization vector is always perpendicular to
the external magnetic field. Therefore, we always assume that when
$\bm{r}_{NV}$ is parallel to $\bm{B}_0$ then $\alpha\to1$ (maximal
spin polarization), while for the other orientations we take
$\alpha=0.7$.

When the angle between the $\bm{r}_{NV}$ and $\bm{B}_0$ vectors is
equal to $\theta_t$, like in the previous case (and for the same
reason), there is a smooth decrease of $\langle\rho_{00}\rangle$
with the field. Besides this, at any orientation of the NV$^-$
center there are three sharp LAC lines at zero field, at 590~G and
at 1024~G. The lines at zero field and at 590~G have different
signs for the different orientations of the NV$^-$ center. It is
worth noting, that the zero-field line has exactly the same
amplitude for the different orientations considered here;
consequently, the total polarization does not have a feature at
zero field. This is consistent with the assumption
\cite{Anishchik2015} that this line is merely due to the
polarization transfer between NV$^-$ centers with different
orientations. In experiments this line becomes visible only
because the luminescence intensity of the differently oriented
NV$^-$ centers is not the same (due to the different absorption
efficiency of the incident light). As a consequence, the
zero-field LAC line (in contrast to all other LAC lines) is a
second-order effect and its intensity quadratically depends on the light
intensity \cite{Anishchik2015}.

Hence, our calculation can account for additional LAC lines at
around 500~G and 590~G, which were reported before and come from
polarization exchange between different defect centers. In the
literature these lines are usually named ``cross-relaxation
lines'' as they are associated with polarization transfer. We
would like to stress that such a term is, perhaps, misleading
because in magnetic resonance cross-relaxation usually means
polarization transfer mediated by stochastic, not dynamics,
processes \cite{Solomon1955,Anderson1962}. In the present case the
``cross-relaxation lines'' obviously originate from coherent
exchange of polarization at corresponding LACs.

\subsection{Comparison with experiment}
Having understood the nature of different lines in LAC spectra, we
can compare our theoretical considerations with the experimental
results and analyze the origin of different lines (or groups of
lines) found experimentally.

\begin{figure}
   \includegraphics[width=0.4\textwidth]{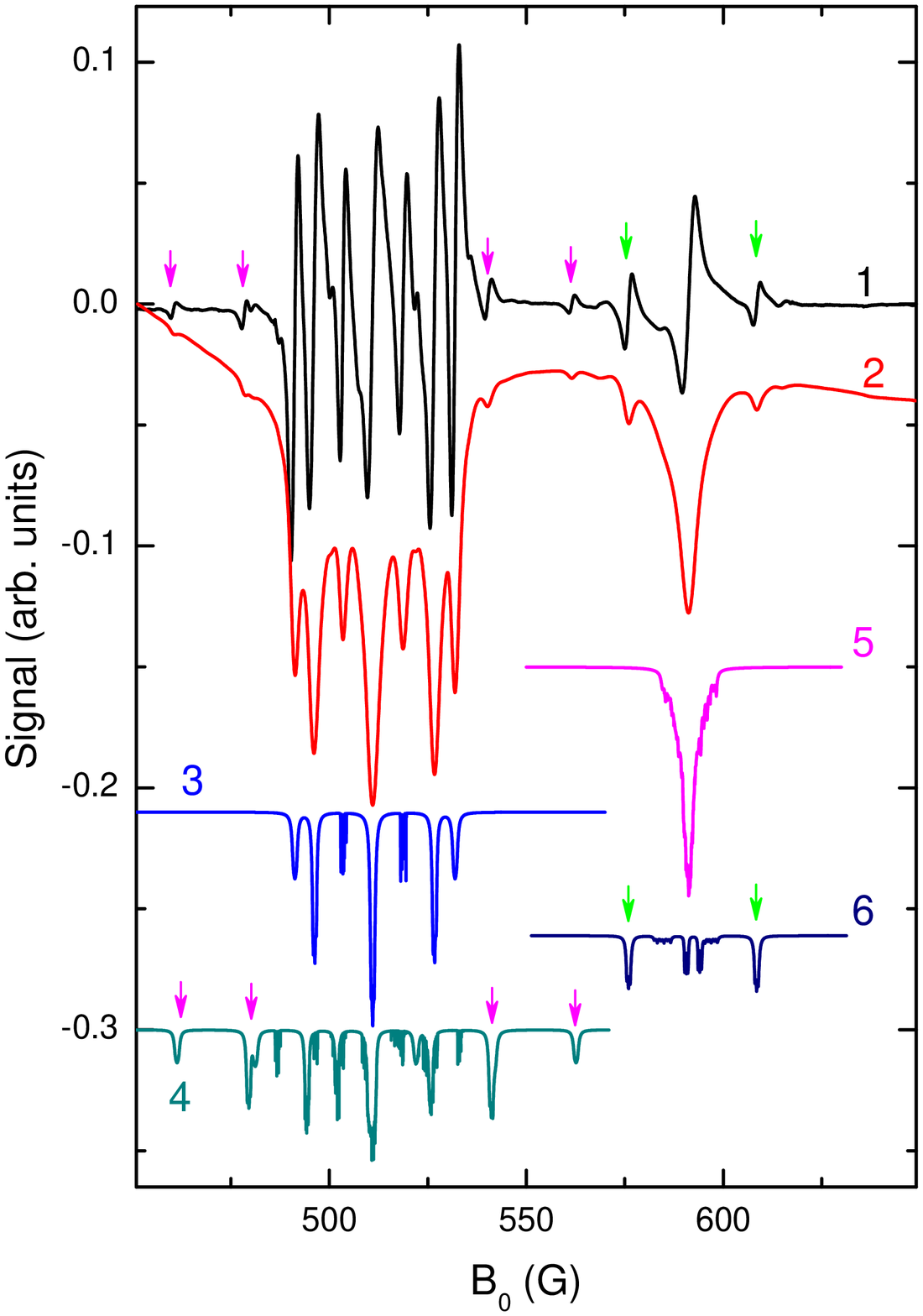}
   \caption{Experimental LAC spectrum in the field range 450-650~G
   (curve 1) and integrated spectrum (curve 2). Curves 3--6 present
   the calculated LAC spectrum for the pair NV$^-$/P1 (curve 3); for
   the pair NV$^-$/P1 assuming that the P1 center has a $^{13}$C
   nucleus at the C$_1$ position (curve 4); for the pair NV$^-$/NV$^-$
   with different orientations
   (curve 5); for the pair NV$^-$/NV$^-$ when one of the centers has
   a $^{13}$C nucleus (curve 6). Arrows indicate the lines in the LAC
   spectrum, which become manifest due to the HFC with $^{13}$C nuclei.
   \label{500}}
\end{figure}

\begin{figure}
   \includegraphics[width=0.4\textwidth]{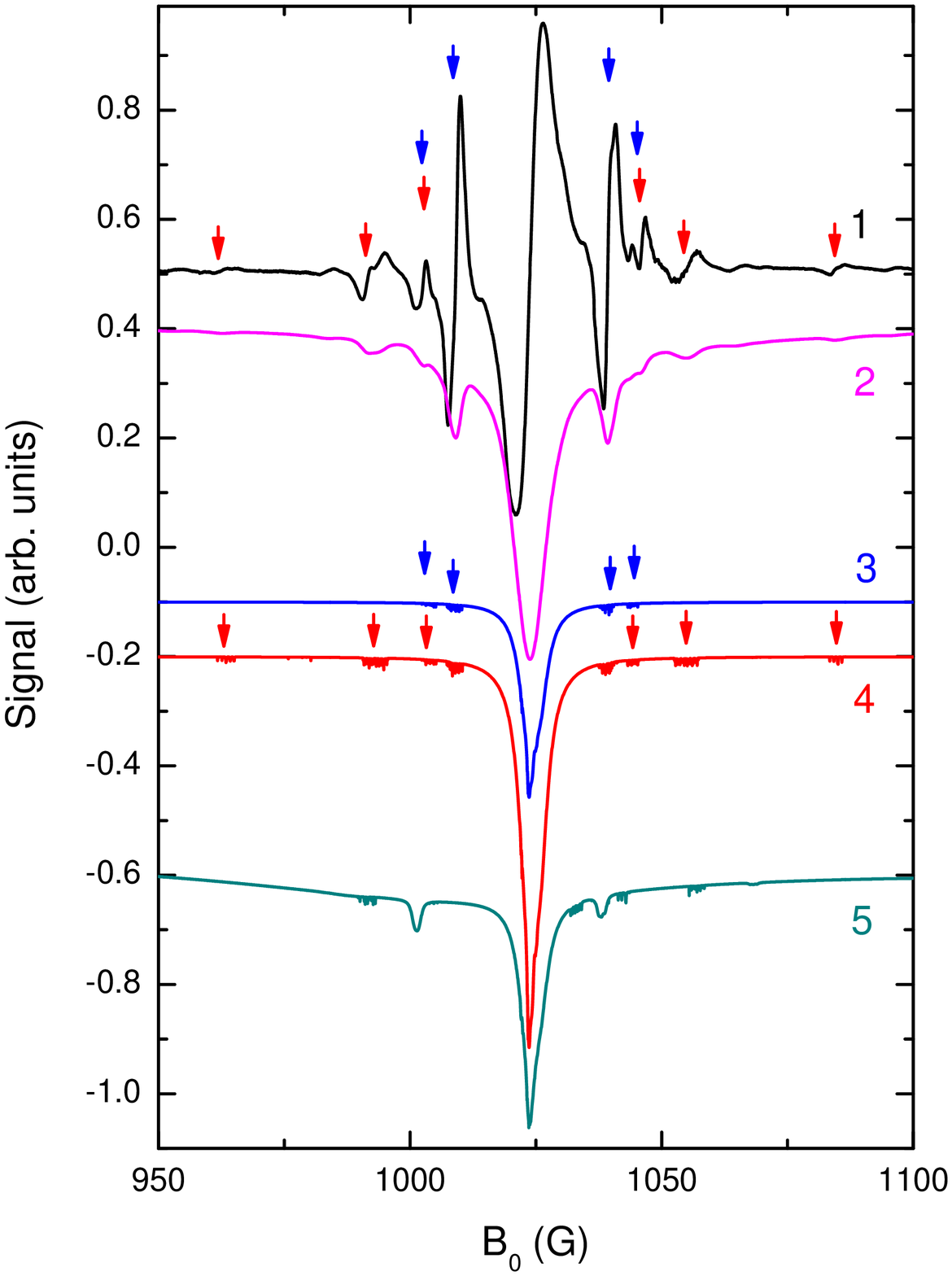}
   \caption{Experimental LAC spectrum in the range 950-1100~G
   (curve 1)  and integrated spectrum (curve 2). Calculation for
   pairs of interacting defect centers:
   pair NV$^-$/P1 (curve 3); pair NV$^-$/P1 taking into account
   coupling of the P1 center to a $^{13}$C spin in the C$_1$
   position (curve 4); pair NV$^-$/NV$^-$ with different orientations
   taking into account a $^{13}$C spin in the proximity of one of the two centers
   (curve 5). Arrows indicate the corresponding lines in the
   experimental and calculated LAC spectra. \label{1000}}\end{figure}

The experimental LAC spectra in their original appearance and also
integrated spectra in the field range 450-650~G are shown in
Fig.~\ref{500}. Additionally, in this figure we present the
calculation results for different pairs of interacting
paramagnetic centers. All parameters of NV$^-$ centers and P1
centers are the same as those given above. In the calculation, the
first center is always an NV$^-$ center oriented parallel to the
$\bm{B}_0$ center; we perform averaging over all four possible
orientations of the P1 center (single orientation parallel to the
field and three orientations where the P1 centers are tilted by
$\theta_t$) and for all three possible orientations of the second
NV$^-$ center (where the angle between $\bm{B}_0$ and
$\bm{r}_{NV}$ is equal to $\theta_t$).

One can see that the calculation performed for the NV$^-$/P1 pair
very well describes the seven previously reported LAC lines in the
range 490-540~G. We would like to stress that our model can
describe not only the positions of the lines, but also their
relative amplitudes. The calculation run for the NV$^-$/NV$^-$
pair also very well reproduces the line at 590~G.

For assigning other lines, namely, the weak satellite lines
indicated in Fig.~\ref{500} we performed a calculation assuming
that the defect centers have the HFC with surrounding carbon
nuclei in the lattice. In the sample studied, $^{13}$C nuclei are
present at the low natural abundance of 1.1\%. For this reason, we
also perform simulations taking into account the fact that one of
the two centers has a $^{13}$C spin in the close proximity to the
paramagnetic defect. For this carbon nucleus in the C1 position we
use the following HFC parameters: for the NV$^-$ center
$A_{||}=A_{\perp}=100$~MHz, for the P1 center $A_{||}=340$~MHz,
$A_{\perp}=140$~MHz \cite{Every1965}.

The LAC spectra calculated taking into account the additional HFC
terms are also presented in Fig.~\ref{500}. In the figure, the arrows
indicate the weak satellite lines in the experimental spectrum and
also in the simulated spectra. One can readily see that there is a
very good agreement between the positions of the lines in the
experimental and calculated spectra. As far as the amplitude of
these lines is concerned, the relative intensities of the lines
caused by the additional HFC in the P1 center are reproduced
properly. Additionally, the calculation can describe the doublet
seen in the spectrum at 477~G.

However, if we keep in mind the low natural abundance of $^{13}$C
nuclei in the sample, the amplitude of the satellite lines in the
calculation is much smaller than that found experimentally. This
is true for the pair NV$^-$/P1 and even to a greater extent for
the pair NV$^-$/NV$^-$. Most likely, this discrepancy is caused by
the effect of field modulation on LAC lines \cite{Anishchik2017}:
at low frequencies the line amplitude strongly increases, this
effect is much more pronounced for weak low-amplitude lines.

One more group of LAC lines is found at around 1024~G; close
inspection shows that in addition to the main line there are also
weak satellite lines seen in the spectrum. In Fig.~\ref{1000} we
show the LAC spectrum in its original appearance together with the
calculated spectra for different pairs of interacting defect
centers: pair NV$^-$/P1 and also pairs NV$^-$/P1 and NV$^-$/NV$^-$
hyperfine-coupled to nearest $^{13}$C nuclei. In the case of the
NV$^-$/P1 pair we take into account the coupling to a $^{13}$C
nucleus in the C$_1$ position in the P1 center. All calculation
parameters are the same as those in previous calculations.

The LAC lines in the spectra calculated for different pairs
strongly overlap, therefore it is problematic to assign specific
lines. We can state only that the weak lines at the
wings of the spectrum shown in Fig.~\ref{1000} are caused by the
HFC to a $^{13}$C nucleus in the P1 center. From the calculation,
we can also deduce that some asymmetry of the spectrum is caused
by the HFCs to carbon nuclei in the NV$^-$/NV$^-$ pair. As fas as
relative amplitude of individual lines are concerned, the
amplitude of the satellite lines is much stronger in experiments
than in the calculation; this effect is even more pronounced as
compared to the spectra shown in Fig.~\ref{500}.

\subsection{Effect of the average evolution time $\tau$}

\begin{figure}
   \includegraphics[width=0.4\textwidth]{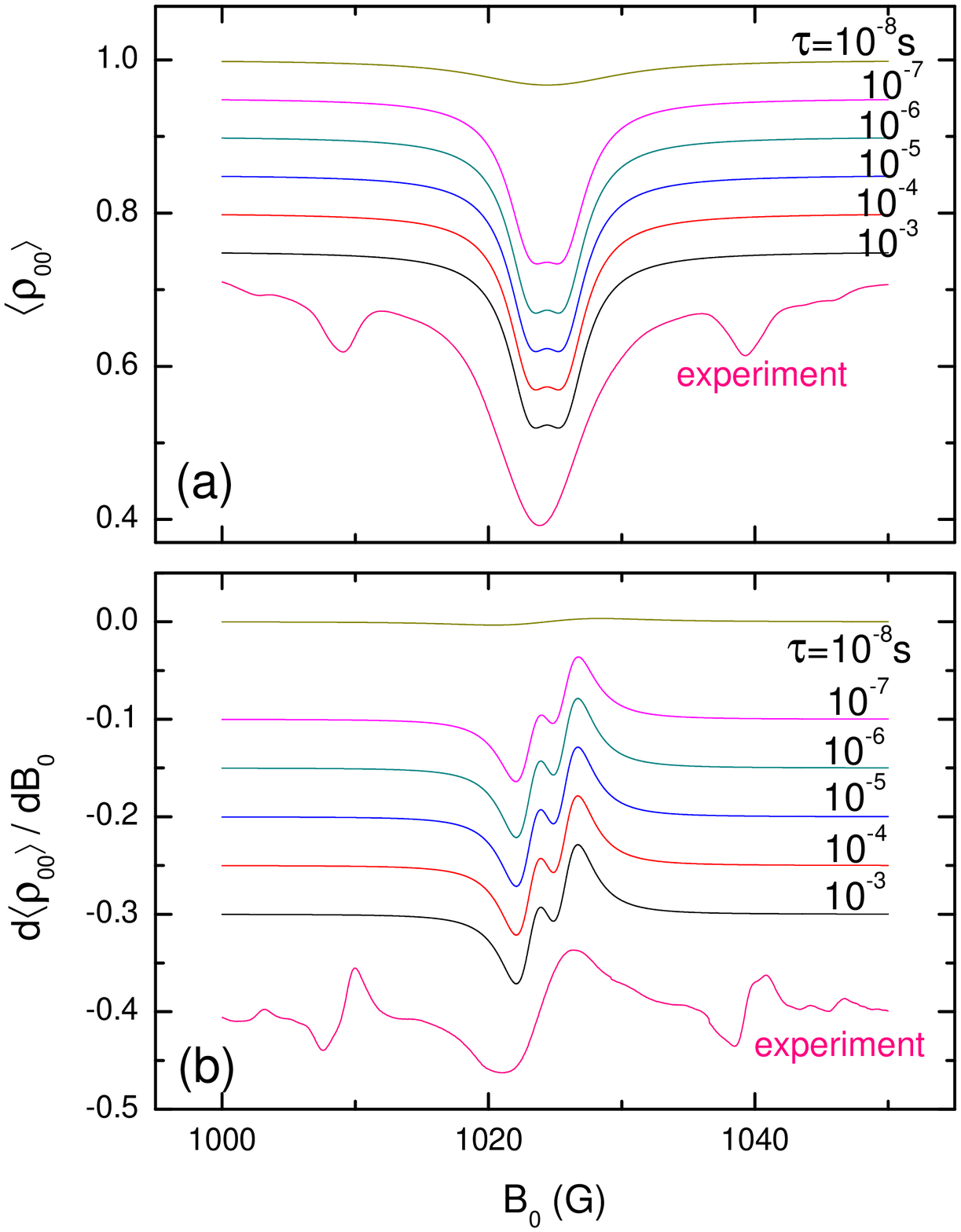}
   \caption{Dependence of the LAC spectrum (shown in the range
   around 1024~G) on the mean evolution time $\tau$ for an
   isolated NV$^-$ center. We present the calculated
   $\langle\rho_{00}\rangle$ field dependence and the integrated
   experimental LAC spectrum (subplot a) and also the $B_0$
   derivative of $\langle\rho_{00}\rangle$ and the experimental
   spectrum in its original appearance (subplot b).
   \label{taudiff}}
\end{figure}

\begin{figure}
   \includegraphics[width=0.4\textwidth]{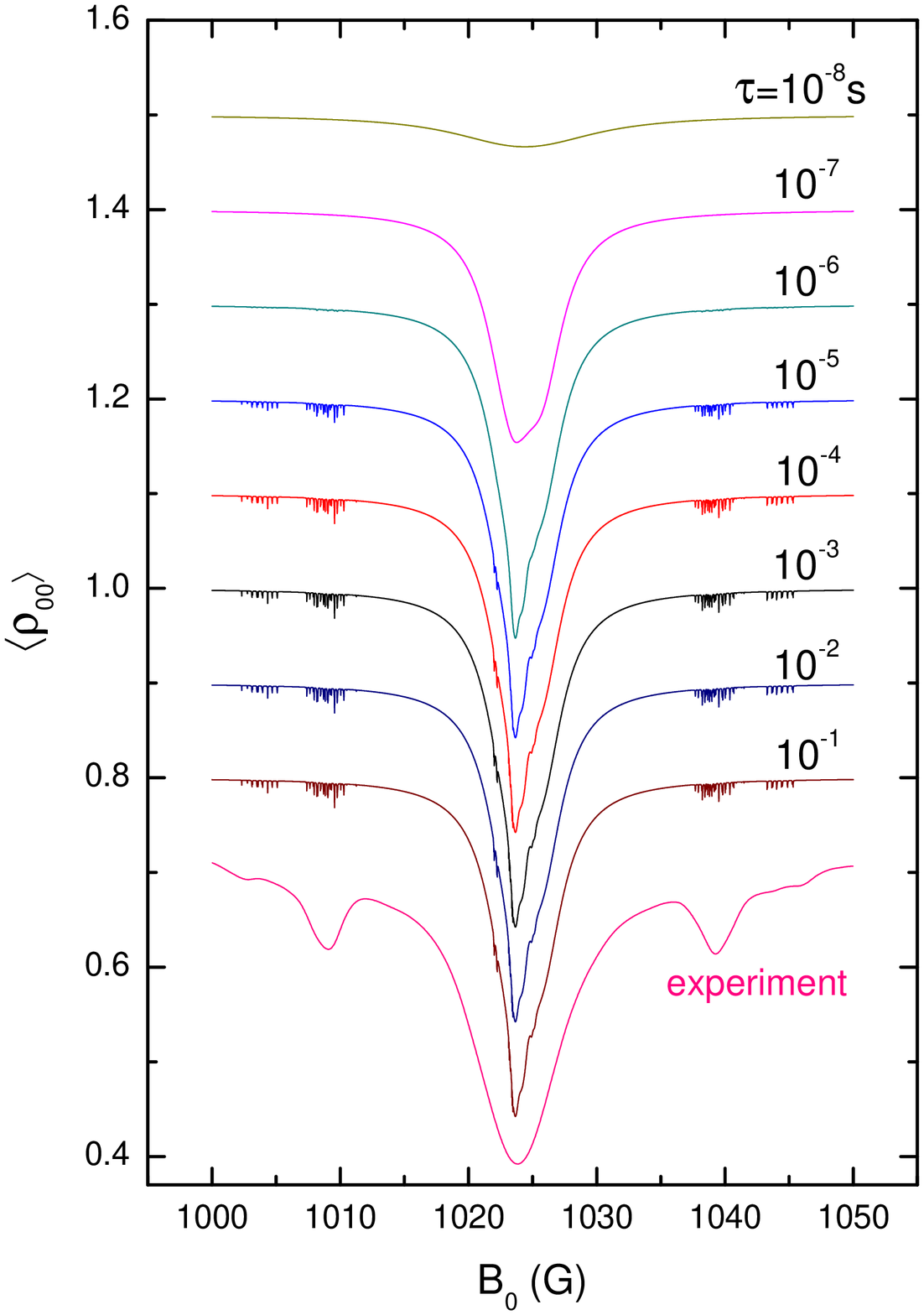}
   \caption{ Dependence of the LAC spectrum for the pair
   NV$^-$/P1 of interacting defect centers on the mean
   evolution time $\tau$ and the experimental LAC spectrum
   (here the integrated spectrum is presented).
   \label{tauP1}}
\end{figure}

Appearance of LAC spectra also depends on the evolution time
$\tau$ because spin mixing at each LAC requires finite time to
develop. Specifically, when the minimal splitting at a LAC is
equal to $2|V_{\mu\nu}|\neq0$ the characteristic mixing time is of
the order of $1/|V_{\mu\nu}|$. When this time is much longer
than $\tau$ the corresponding LAC would not reveal itself as a
feature in the magnetic field dependence of photoluminescence.
Hence, to observe a feature coming from a LAC the evolution time
should be greater than or comparable to $1/|V_{\mu\nu}|\neq0$;
$\tau\approx1/|V_{\mu\nu}|\neq0$ gives a threshold value for
observing the feature.

In our calculation method, the evolution time $\tau$ is an input
parameter, which is difficult to estimate. For this reason, we
need to know how critical is the $\tau$ dependence of LAC line
intensities. In Figs.~\ref{taudiff}, \ref{tauP1} we present the
LAC spectra calculated for different $\tau$ values considering a
pair of defects coupled by dipole-dipole interaction of the
strength of 1~MHz. The parameter $\tau$ is varied over a wide
range from $10^{-8}$ to $10^{-1}$~s. One can readily see that
there is a certain threshold $\tau$ value: above this value the
spectrum almost does not change. For the LAC line at 1024~G this
value is approximately $10^{-7}$~s  (see Fig.~\ref{taudiff}). The
weak satellite lines are more sensitive to the $\tau$ value, as
can be seen from Fig.~\ref{tauP1}. For these lines, the threshold
$\tau$ value is about $10^{-5}$~s. In all calculations, we use the
value $\tau=10^{-4}$~s, which guarantees correct evaluation of the
spectra in all cases (meaning that all LAC lines of interest
become manifest).

The theoretical curves in Fig.~\ref{taudiff} exhibit a splitting
of the main LAC line at 1024~G. In experiments such an effect is
not observed, which we attribute to a relatively high amplitude of
field modulation, which is 0.5~G, lowering the resolution. From
comparison of Figs.~\ref{taudiff} and \ref{tauP1} we can draw one
more conclusion: in Fig.~\ref{taudiff} the position of the LAC
line of an isolated NV$^-$ center is shifted to higher fields as
compared to the experimental observation (by approximately 0.5~G).
For the NV$^-$/P1 pair, this line is slightly distorted (see Fig.~
\ref{tauP1}) so that the low-field component becomes stronger and
the high-field component becomes weaker. As a result, there is
only one LAC line present in the spectrum and its position
precisely fits to that found experimentally. Hence, we can
conclude that the pair NV$^-$/P1 makes the dominant contribution
to the line at 1024~G.

\section{Discussion}

Hence, we present a method, which can be used to treat
quantitatively LAC spectra in diamond crystals containing
paramagnetic color centers. The method uses only standard linear
algebra methods, such as matrix multiplication and solution of the
eigen-problem of a hermitian matrix. Consequently, we are able to
treat multi-spin systems that are described by Hamiltonians of a
large dimensionality. For instance, here we consider the system of
four spins 1 (two spin-1 paramagnetic defects each having a spin-1
$^{14}$N nucleus) and one spin-1/2 $^{13}$C nucleus; in this case
the Hamiltonian is a $162\times162$ matrix. Our approach can be
easily extended to systems containing color centers other than
NV$^-$ centers. The main simplifying assumption used here is
neglecting the spin evolution in the excited state: when
necessary, the spin evolution in the excited state can be treated
by using the same method by redefining the time $\tau$ as the
lifetime of the excited state. A more rigorous approach, which
would explicitly take into account the evolution in both states
and transitions between them can also be implemented, see e.g.
Ref. \cite{Sosnovsky2018}, but it is significantly more time
consuming.

Our treatment shows that the positions of LAC lines can be
accurately described by the proposed theoretical approach. We
would like to note that the calculations shown here do not use any
free parameters other than the evolution time (which does not
strongly affect LAC spectra): the parameters of the relevant
interaction tensors are taken from independent measurements
reported in literature. As far as the relative amplitudes of LAC
lines are concerned, our method precisely describes the amplitudes
of lines belonging to the same group. This is true, for instance,
for the lines found in the range 490-540~G, which are conditioned
by coupling of the electron spins of NV$^-$ centers and P1
centers. The same situation holds for the lines found in the range
450-650~G: these lines are due to $^{13}$C nuclei in the C$_1$
position of the P1 center. However, the satellite lines at around
1024~G, as well as the satellite lines in the range 450-650~G,
which are due to HFC to $^{13}$C nuclei (present only in low
concentration), in experiments are much stronger than expected
from the theoretical treatment.

Field modulation effects are not included in our simulations (this
is not feasible with available computational resources) but on the
qualitative level the influence of modulation can be rationalized
from our previous work \cite{Anishchik2017}. Specifically, we have
considered the effect of field modulation on the amplitude of LAC
lines and demonstrated that the line amplitudes strongly increase
at low modulation frequencies. Furthermore, for weak lines such an
increase is associated with polarization transfer to nuclear spins
giving rise to an even stronger increase of the line amplitude.
Specifically, for weak LAC lines the increase in amplitude upon
decreasing the modulation frequency from 12.5~kHz to 17~Hz is an
order of magnitude stronger than for intense lines, e.g., than for
the LAC line at 1024~G. Thus, the amplitude ratio calculated by
using our method is more similar to experimental observations at
the modulation frequency of 12.5~kHz, rather than to those at
17~Hz. Numerical calculations \cite{Anishchik2017} show that such
an enhancement of LAC lines is due to the fact that nuclear spins,
which relax much slower than the electron spins, can store spin
polarization. After many excitation-radiation cycles (or
excitation-radiationless relaxation), the polarization is
transferred from NV$^-$ centers to nuclei. Due to the much slower
spin relaxation, nuclei can store polarization; subsequently,
polarization transfer back to NV$^-$ can occur. Such polarization
transfer is most efficient at low modulation frequencies; in the
present theoretical model it is not taken into account.

\section{Conclusions}

We propose a numerically efficient method for describing LAC
spectra of NV$^-$ centers in diamond crystals. From the
mathematical viewpoint, the method makes use of the solution of
the eigen-problem of hermitian matrices and basic matrix
operators. Standard numerical algorithms for such calculations are
well established and efficient, allowing one to implement the
proposed method.  The method allows one to predict not only the
positions of LAC lines, but also their shapes and relative
intensities; hence, we are able to model the main groups of lines
in LAC spectra, assign lines to specific LACs and analyze
satellites coming from $^{13}$C nuclei at low natural abundance.
We expect that the proposed method can be used to determine
magnetic resonance parameters of dark defect centers interacting
with NV$^-$ centers and also to investigate such interactions.

\begin{acknowledgments}

The work was supported by the Russian Foundation for Basic
Research (Grant No. 16-03-00672 and 17-03-00932).

\end{acknowledgments}

\bibliography{Anishchik}

\end{document}